
       \def\A{\cal A}
       \def\AG{{\cal A}/{\cal G}}
       \def\AGbar{\overline{\AG}}
       \def\HA{{\cal HA}}
       \def\HG{\cal HG}
       
       \def\S{\Sigma}
       \def\C{$C^\star$-algebra}
       \def\HA{\cal HA}
       \def\Ab{\bar{A}}
       \def\a{\alpha}
       \def\at{\tilde{\a}}
       \def\b{\beta}
       \def\bt{\tilde{\b}}
       \def\L{{\cal L}_{x_o}}
       \def\SU{[SU(2)]^n/Ad}


       \def\bra#1{\langle\, #1\, |}
 
       \def\d{\partial}
       \def\half{{\textstyle{1\over2}}}
       \def\implies{\Rightarrow}
       \def\IP#1#2{\langle\, #1\, |\, #2\, \rangle}
       
       \def\lint{\int\nolimits}

       \def\ovr{\overline}
       \def\real{{\rm I\!R}}
       \def\to{\rightarrow}
       \def\Tr{\mathop{\rm Tr}}
       \def\tw{\widetilde}

\magnification=1200

\centerline{{\bf Quantum Gravity: A Mathematical Physics Perspective}}
\bigskip
\centerline{Abhay Ashtekar}
\centerline{\it Center for Gravitational Physics and Geometry}
\centerline{\it Penn State University, University park, PA16802-6300}
\bigskip
\bigskip

{\bf 1. Introduction}

The problem of quantum gravity is an old one and over the course of
time several distinct lines of thought have evolved. However, for
several decades, there was very little communication between the two
main communities in this area: particle physicists and gravitation
theorists. Indeed, there was a lack of agreement on even what the key
problems are. By and large, particle physics approaches focused on
perturbative techniques.  The space-time metric was split into two
parts: $g_{\mu\nu}=\eta_{\mu\nu} + G h_{\mu\nu}$, $\eta_{\mu\nu}$
being regarded as a flat kinematic piece, $h_{\mu\nu}$ being assigned
the role of the dynamical variable and Newton's constant $G$ playing
the role of the coupling constant. The field $h_{\mu\nu}$ was then
quantized on the $\eta_{\mu\nu}$-background and perturbative
techniques that had been so successful in quantum electrodynamics were
applied to the Einstein-Hilbert action. The key problems then were
those of handling the infinities. The gravity community, on the other
hand, felt that a central lesson of general relativity is that the
space-time metric plays a dual role: it is important that one and the
same mathematical object determine geometry {\it and} encode the
physical gravitational field. From this perspective, an ad-hoc split
of the metric goes against the very spirit of the theory and must be
avoided. If one does not carry out the split, however, a theory of
quantum gravity would be simultaneously a theory of quantum geometry
and the notion of quantum geometry raises a variety of conceptual
difficulties. If there is no background space-time geometry --but only
a probability amplitude for various possibilities-- how does one do
physics? What does causality mean?  What is time? What does dynamics
mean? Gravity theorists focused on such conceptual issues.  To
simplify mathematics, they often truncated the theory by imposing
various symmetry conditions and thus avoided the field theoretic
difficulties.  Technically, the emphasis was on geometry rather than
functional analysis. It is not that each community was completely
unaware of the work of the other (although, by and large, neither had
fully absorbed what the other side was saying). Rather, each side had
its list of central problems and believed that once these issues were
resolved, the remaining ones could be handled without much
difficulty. To high energy theorists, the conceptual problems of
relativists were perhaps analogous to the issues in foundations of
quantum mechanics which they considered to be ``unimportant for real
physical predictions.'' To relativists, the field theoretic
difficulties of high energy physicists were technicalities which could
be sorted out after the conceptual issues had been resolved. This is
of course a simplified picture. My aim is only to provide an
impression of the general state of affairs%
\footnote{.}{To appear in the proceedings of the workshop on {\sl
Mathematical Physics Towards the XXIst Century}, held at Beer-Sheva,
Israel in March 1993.}

Over the last decade, however, there has been a certain rapprochement
of ideas on quantum gravity. Each side has become increasingly aware
of the difficulties that were emphasized by the other. By and large,
high energy theorists now agree that the non-perturbative techniques
are critical. They see that the underlying diffeomorphism invariance
should be respected. Even if one introduces a background structure
for, e.g., regularization of operators, the final result should not
make reference to such structures (unless of course they are
physically important). Relativists have come to recognize that the
field theoretic divergences have to be faced squarely. There is now a
general agreement that although the truncated theories are interesting
and provide insights into the conceptual (and certain mathematical)
problems faced by the full theory, they are essentially toy models
whose value, in the final analysis, is quite limited since they have
only a finite number of degrees of freedom.  Thus, the sets of goals
of the two communities have moved closer.

These recognitions do not imply, however, that there is a general
consensus on {\it how} all these problems are to be resolved.  Thus,
there are again many approaches. But this diversity in the lines of
attack is very healthy.  In a problem like quantum gravity, where
directly relevant experimental data is scarce, it would be an error if
everyone followed the same path. As Feynman (1965) put it:
{\smallskip\narrower{\sl \noindent It is very important that
we do not all follow the same fashion. ... It's necessary to increase
the amount of variety ... and the only way to do is to implore you few
guys to take a risk with your lives that you will not be heard of
again, and go off in the wild blue yonder to see if you can figure it
out.}\smallskip}
\noindent And quite a few groups have taken the spirit of this advice
seriously and ``gone off in the yonder to figure it out.'' What is
striking is that, in spite of the diversity of their methods, some
their results are qualitatively similar. One message that seems to
keep coming back is that not only can one not assume a flat
Minkowskian geometry at the Planck scale but in fact even the more
general notions from Riemannian geometry would fail. The continuum
picture itself is likely to break down. The lesson comes from certain
computer simulations of 4-dimensional Euclidean gravity, (see e.g.
Agishtein \& Migdal (1992)) from string theory (see e.g. Gross \&
Mende (1988), Amati et al (1990), Aspinwall (1993)) and from canonical
quantization of 4-dimensional general relativity (Ashtekar et al
(1992)). The detailed pictures of the micro-structure of space-time
that arises in these approaches is quite different at least at first
sight and it is not clear that these pictures can be reconciled with
one another in detail. Nonetheless, there {\it are} certain
similarities in the results and most of them are obtained by using
genuinely non-perturbative techniques.

The purpose of this article is to give a flavor of these ideas and
techniques to mathematical physicists.  I should emphasize that this is
not a systematic survey. In particular, I will concentrate just on one
approach --non-perturbative, canonical treatment of 4-dimensional,
Lorentzian general relativity. Even within this approach, there are
over 350 papers and I can not do justice to them in this limited
space. (For detailed reviews, see, e.g., Ashtekar (1991,1992) and
Pullin (1993).) Rather, I will just present a few results that may be
of interest to this audience, indicating, wherever possible, the
degree of precision and rigor of the underlying calculations.

However, since the goal of the conference is to look to future,
``towards the 21st century,'' and since the organizers asked us to try
to ``inspire rather than merely inform,'' I will begin in section 2
with a few remarks that may perhaps seem provocative to some
mathematical physicists.  In doing so, however, I am following the
lead of other speakers and my hope is the same as theirs: these
remarks may lead to stimulating exchanges of ideas and perhaps a
re-examination of some of the basic premises.  In section 3, the main
difficulties of quantum gravity are outlined from the perspective of
mathematical physics.  Section 4 summarizes some recent developments.
While general relativity is normally regarded as a theory of
metrics, it can be recast as a dynamical theory of connections
(Ashtekar, 1987). This shift of emphasis has two important
consequences. First, it brings general relativity closer to theories
of other interactions and one can draw on the numerous techniques that
have been developed to quantize these theories. Second, the shift
simplifies the basic equations considerably making them low order
polynomials in the basic variables. (They are non-polynomial in the
metric variables.)  Recently, a number of rigorous results have been
obtained to analyze theories of connections, particularly the ones
such as general relativity in which there is an underlying
diffeomorphism invariance.  The main idea here is to use algebraic
methods to develop an integration theory on the space $\AG$ of
connections modulo gauge transformations and to explicitly construct a
measure which is diffeomorphism invariant. These results now provide
the foundation for a non-perturbative approach to quantum gravity
based on Hamiltonian methods. They may also have other applications in
mathematical physics.  In section 5, this framework is used to show
the existence of states in the full, non-perturbative quantum gravity
which approximate a classical metric when coarse-grained at scales
much larger than the Planck length {\it and} which exhibit a specific
discrete structure at the Planck scale.  If such a state is used in
place of Minkowski space-time, the ultra-violet difficulties of
Minkowskian field theories may disappear altogether. This is a
concrete illustration of the results in non-perturbative quantum
gravity that may have an impact on quantum field theory.

\bigskip\bigskip
\goodbreak
{\bf 2. The ultraviolet catastrophe: A matter of gravity?}

As we all know, for over 40 years, quantum field theory has been in a
somewhat peculiar situation as far as realistic models are considered.
On the one hand, perturbative treatments are available for, say, the
electro-weak interaction and the results are in excellent agreement
with experiments. It is clear therefore that there is something
``essentially right'' about these theories. On the other hand, their
mathematical status has continued to be dubious and it has not been
possible to say precisely what is right with them. And this is not
because of lack of effort. Already by early sixties, quantum field
theory had become an intellectually coherent subject (recall that {\it
PCT spin and statistics and all that} was published in 1964). A
considerable amount of imaginative --and often heroic-- effort has
gone in to the field since then. And yet none of the physically realistic
quantum field theories has reached a sound status in 4 space-time
dimensions.

I think it is fair to say that the general attitude in the
mathematical physics community is that the difficulties are of a
mathematical nature.  Realistic 4-dimensional theories are extremely
involved; as Professor Wightman put it at the conference, ``handling
them with the present techniques is hellishly complicated.''  So, the
overall feeling seems to be that there is no obstacle of principle to
construct a non-Abelian gauge theory such as QCD in 4-dimensions, but
new mathematical tools are needed to make the task practicable.

One might worry about the fact that in all these theories, one uses
Minkowski space as the underlying space-time, thereby ignoring all the
Planck scale effects. Could this be a source of the difficulty? The
general belief in the community seems to be that this is not the case;
no new physical input from the Planck scale is needed to rigorously
construct the quantum theory that underlies the standard model of
particle physics. (This view was, for example, expressed in Professor
Buchholtz's talk.)  Sure, the theory would be an approximation in that
one would not be able to trust its predictions below, say $10^{-17}$cm.
But it would be internally consistent and agree with Nature as far as
laboratory physics is concerned. Sure, we are idealizing the
space-time geometry. But one does this all the time and such
idealizations always work: our general experience in physics tells us
that, somehow, the phenomena at different scales decouple
approximately. Indeed, very little progress could have occurred in
absence of this decoupling.  After all, when engineers build bridges,
they don't have to worry about the fact that it is quantum mechanics
that governs the atoms of the bridge; they just use classical,
Newtonian physics. They succeed because there is a factor of $10^{12}$
between the scale of the bridge and the atomic scale.  There is a
factor of $10^{16}$ between (say) the weak-interaction scale and the
Planck length! Surely, one says, the Planck-size effects are
unimportant to the problem of constructing a consistent (and therefore
in particular, finite) quantum theory underlying the standard model.

I would like to argue that this is not necessarily the case. Recall,
first, that the key difficulties of quantum field theory are the
ultra-violet divergences. These arise {\it precisely} because we allow
virtual processes involving arbitrary number of loops, each carrying
an integral over arbitrarily large momenta.  And it is not {\it our}
choice to allow or disallow them: we are forced into it by the general
principles of quantum field theory which include Poincar\'e
invariance. Surely, as the energy involved becomes bigger, the
approximation that one can ignore the gravitational effects becomes
worse. If the microstructure of space-time is qualitatively different
from that given by the continuum picture, the whole procedure is
flawed; we shouldn't --and couldn't-- integrate to arbitrarily high
momenta since that is equivalent to integrating to arbitrarily small
distances. Let us consider an analogy with atomic physics where we can
successfully use non-relativistic quantum mechanics. Suppose for a
moment that there was a general requirement that arose from the
quantum principles which forced us, in the calculation of, say, the
ground state energy of the hydrogen atom, to consider electrons with
velocities arbitrarily close to the speed of light. {\it Then}, had we
ignored special relativity altogether, we {\it would} probably have
got an inconsistent theory: The hypothetical quantum principle would
have {\it forced} us to bring special relativity into the treatment.
The standard treatment of atomic physics in non-relativistic quantum
mechanics is internally consistent {\it because} the calculation
scheme does not involve steps which violate the basic premise and
limitation of non-relativistic quantum mechanics.  (It agrees well
with experiments because, in addition, these approximations are met in
Nature.) Let us return to quantum field theory. In spite of the fact
that we are interested here in processes in which physical energies
(and masses of particles) are small compared to the Planck mass, we
are forced to allow {\it virtual} processes involving arbitrarily high
energies and these do probe the Planck-scale structure of the
space-time geometry. But we insist on ignoring this structure
altogether. {\it That}, it may well be, is the {\it physical} source
of the ultraviolet catastrophe; it may be a matter of gravity.

Indeed, there are other instances in physics where mathematical
difficulties signalled the need for changing the basic physical
premise. Consider for instance the action at a distance models in
classical, relativistic physics%
\footnote{${}^\dagger$}{This example was suggested to me by Jose
Mour\~ao}.
The system of integro-differential equations one obtains is hard to
manage mathematically and, if taken seriously, raises questions of
predictibility in relativistic physics.  The ``origin'' of these
difficulties lies in the physical inadequacy of the basic assumptions.
Once we bring in the field degrees of freedom, these mathematical
problems go away. Now, classical physics is described by a system of
hyperbolic differential equations and causality is manifest. Another
example is the birth of quantum mechanics itself. The mathematical
description of the black-body radiation broke down in the framework of
classical physics and pointed to the necessity of a radical revision
of that framework. There are a number of other examples. Indeed, most
radical changes of the conceptual framework are inspired, at least in
part, by the fact that the older framework ran into serious
mathematical difficulties. This does not of course imply that the same
{\it must} happen with the ultra-violaet divergences. These are only
analogies. And there is no irrefutable evidence that more
sophisticated mathematical techniques will not suffice to construct a
consistent quantum field theory for the standard model.  However, the
examples suggest that we should be open to such a possibility.

I will conclude with a related but somewhat different point. Let me
again use an analogy. Suppose, for a moment that special relativity
had not been discovered. One might have learnt painfully that the
predictions of Newtonian mechanics are not quite correct and have to
be supplemented by powers of $v/c$ where $v$ is the velocity of the
object under consideration. If we had been sufficiently clever, we
would have discovered that to compute any physical effect, one can use
a perturbation series in the powers of $v/c$. One could get
sophisticated and worry about whether such series actually converge.
This is perhaps similar to the present situation with the perturbation
theory for the electro-weak interaction. In the hypothetical case of
``special relativity'', proving convergence and other mathematical
properties of the resulting series would be instructive. Like Lorentz
and Poincar\'e, one could have even discovered the Lorentz
transformations and found all sorts of equations which are ``true.''
However, without the {\it physical} shift of scenario that was
provided to us by Einstein --that there is no absolute simultaneity--
we would still be missing key insights. In a real sense, we would not
really ``understand'' what the equations were telling us. The situation
could be similar with the standard model. Suppose we do succeed in
giving a mathematical meaning to the perturbative results. We will
have a nice, consistent theory with convergence proofs. But it is
possible that we may still be missing some key insights because we
ignored the Planck scale physics. Indeed, it is not obvious that {\it
all} effects of the quantum nature of geometry will be confined to the
Planck scale. Let us take special relativity. It is true that most
effects are corrections in powers of $v/c$ to the predictions of
non-relativistic physics. But now and again, there are also {\it
qualitative} predictions that have nothing to do with how large the
velocity $v$ of the particles involved is. Conversion of mass into
enormous amount of energy happens in nuclear processes where all
velocities are small. There is a prediction that associated with every
particle there is an anti-particle. There is the CPT theorem. These are
all qualitative effects completely unrelated to how fast the particles
in question are moving. They come about because special relativity
shifts the very paradigm within which one operates. The entire
mathematical framework of quantum physics changes abruptly. The
problems change. The tools change.  The concepts change. There is a
possibility that quantum field theory may undergo a similar radical
change once the concepts from quantum gravity are brought in. Indeed,
we will see in the subsequent sections that quantum gravity does make
strong demands on how one should formulate and analyze problems.  It
insists on diffeomorphism invariance whence there is no background
metric or causal structure; it trivializes the Hamiltonian and puts
the burden of dynamics on the constraints of the theory; it introduces
an essential non-locality through physical observables. The ground
rules therefore seem quite different from those we are used to in
Minkowskian quantum field theories.

Mathematical physicists can raise an immediate objection against the
all these possibilities. After all, there do exist well-defined,
consistent quantum field theories in 2 and 3 space-time dimensions.
Why don't the Planck scale problems raise their annoying head there?%
\footnote{${}^\dagger$}
{Indeed, Professor Chayes did ask this question during my
talk! }
It turns out that there is a dramatic change in the properties of the
gravitational field starting precisely at 4 dimensions! It acquires
its {\it local} degrees of freedom only in 4 and higher dimensions.
In dimension 3 or lower, there are no gravity waves and no gravitons.
Therefore, the ultraviolet problems of field theories are, in a sense,
decoupled from the quantum gravitational field. Indeed, one can turn
the argument around and ask if it is a pure coincidence that the
number of dimensions for which the ultra-violet problems of field
theories seem so difficult to handle happens to be precisely the one
at which gravity comes on its own. Or, is there a lesson lurking here
that we have ignored?

I want to emphasize again that I do not regard the arguments given in
this section as conclusive. It is a viable, logical possibility that a
consistent quantum field theory incorporating the standard model will
exist in 4-dimensions and will contain {\it all} the physics that is
relevant to the scale of these interactions. Quantum gravity may
have no effect whatsoever on this physics.  However, I feel that this
``mainstream'' viewpoint in mathematical physics is also not
watertight. There {\it are} differences between the current situation
in field theory and previous examples in the history of physics where
a clean decoupling occurred between physics at one scale and that at
another. And, in the interest of variety, I believe it is important
not to ignore altogether the possibility that the ultra-violet
catastrophe may, in the end, be a matter of gravity.

\bigskip\bigskip
\goodbreak
{\bf 3. Difficulties of quantum gravity}

The importance of the problem of unification of general relativity and
quantum theory was recognized as early as 1930's and a great deal of
effort has been devoted to it in the last three decades.  Yet, we seem
to be far from seeing the light at the end of the tunnel.  Why is the
problem so hard? What are the principal difficulties? Why can we not
apply the quantization techniques that have been successful in
theories of other interactions? In this section, I will address these
questions {\it from the perspective of mathematical physics}.

The main difficulty, of course, is the lack of experimental data with
a direct bearing on quantum gravity. One can argue that this need not
be an unsurmountable obstacle. After all, one hardly had any
experimental data with a direct bearing on general relativity when the
theory was invented.  Furthermore, the main motivation came from the
incompatibility of Newtonian gravity with special relativity. We face
a similar situation; we too are driven by what appears to be a fundamental
tension between general relativity and quantum theory.
However, it is also clear that the situation with discovery of general
relativity is an anomaly rather than a rule. Most new physical
theories --including quantum mechanics -- arose and were continually
guided and shaped by experimental input. In quantum gravity, we are
trying to make a jump by some twenty orders of magnitude --from a
fermi to a Planck length. The hope that there is no dramatically new
physics in the intermediate range is probably just that --a hope.

The experimental status, however, makes the situation even more
puzzling. If there is hardly any experimental data, theorists should
have a ball; without these ``external, bothersome constraints,'' they
should be able to churn out a theory a week. Why then do we not have a
{\it single} theory in spite of all this work? The brief answer, I
think, is that it is {\it very} difficult to do quantum physics in
absence of a background space-time. We have {\it very} little
experience in constructing physically realistic, diffeomorphism
invariant field theories. Indeed, until recently, there were just a
handful of examples, obtained by truncating general relativity in
various ways. It is only in the last three years or so that a
significant number of diffeomorphism invariant models with an infinite
number of degrees of freedom has become available, still, however, in
low space-time dimensions.

As mentioned in the Introduction, one way out of this quandary --tried
by the high energy physicists-- was to simply break the diffeomorphism
invariance to start with and introduce a flat background metric. As is
well-known, however, the resulting perturbative quantum field theory
is non-renormalizable. In the high energy community, this was
considered a fatal flaw.  At first, it was thought that the problem is
with the starting point --general relativity. Therefore, attempts were
made to modify Einstein's theory.  Perhaps the most notable of these
modifications were the higher derivative theories and supergravity.
However, these attempts at defining a local quantum field theory for
gravity (with matter) which is consistent order by order in the
perturbation expansion failed.  Finally, these developments led, in
the mid-eighties, to string theory. Since there were several talks on
this subject in the conference, I will restrict myself just to a one
line summary here: The perturbation series in string theory is
believed to be finite order by order (in the string tension) but the
series is believed not to be even Borel-summable. As a result, a great
deal of effort is being devoted to constructing the theory
non-perturbatively.

Returning to quantum general relativity, the failure of perturbation
theory would, presumably, not upset the mathematical physicists a
great deal. After all, they know that in spite of perturbative
non-renormalizability, a quantum field theory {\it can} exist
non-perturbatively. This point was discussed in some detail in
Professor Klauder's talk and Professor Wightman commented on it in the
context of the Gross-Neveu model in 3-dimensions. Indeed, in the case
of (GN)${}_3$, there appears to be no fundamental difference from
renormalizable models. In particular, we learnt in this conference
that the conjecture that such models should be based on distributions
which are worse behaved than tempered distributions has been shown to
be false. So, at a basic mathematical level, non-renormalizability
seems not be a fundamental consideration.  Returning to gravity, as I
will indicate below, there is some evidence from numerical simulations
that quantum general relativity itself may exist non-perturbatively.
One might therefore wonder: why have the standard methods developed by
mathematical physicists not been applied to the problem of quantum
gravity? What are the obstacles? Let me therefore consider some
natural strategies that one may be tempted to try and indicate the
type of difficulties one encounters.

First, one might imagine {\it defining} the goal properly by writing
down a set of axioms. In Minkowskian field theories the Wightman and
the Haag-Kastler axioms serve this purpose. Can we write down an
analogous system for quantum gravity, thereby spelling out the goals
in a clean fashion? Problems arise right away because both systems
of axioms are rooted in the geometry of Minkowski space and in the
associated Poincar\'e group. Let me consider the Wightman system
(Streater \& Wightman 1964) for concreteness. The zeroth axiom asks
that the Hilbert space of states carry an unitary representation of
the Poincar\'e group and that the 4-momentum operator have a spectrum
in the future cone; the second axiom states how the field operators
should transform; the third axiom introduces micro-causality, i.e.,
the condition that field operators should commute at space-like
separations; and, the fourth and the last axiom requires asymptotic
completeness, i.e., that the Hilbert spaces ${\cal H}^\pm$ of
asymptotic states be isomorphic with the total Hilbert space. Thus,
four of the five axioms derive their meaning directly from Minkowskian
structure. It seems quite difficult to extend the zeroth and the
fourth axioms already to quantum field theory in topologically
non-trivial space-times, leave alone to the context in which there is
no background metric what so ever. And if we just drop these axioms,
we are of course left with a framework that is too loose to be useful.

The situation is similar with the Osterwalder-Schrader system. One
might imagine foregoing the use of specific axioms and just using
techniques from Euclidean quantum field theory to construct a suitable
mathematical framework.  This is the view recently adapted by some
groups using computer simulations.  These methods have had a great
deal of success in certain exactly soluble 2-dimensional models. The
techniques involve dynamical triangulations and have been extended to
the Einstein theory in 4 dimensions (see, e.g., Agishtein \& Migdal
(1992)). Furthermore, there is some numerical evidence that there is a
critical point in the 2-dimensional parameter space spanned by
Newton's constant and the cosmological constant, suggesting that the
continuum limit of the theory may well exist. This is an exciting
development and interesting results have now been obtained by several
groups. Let us be optimistic and suppose that a well-defined Euclidean
quantum theory of gravity can actually be constructed. This would be a
major achievement. Unfortunately, it wouldn't quite solve the problem
at hand. The main obstacle is that, as of now, there is no obvious way
to pass from the Euclidean to the Lorentzian regime! The standard
strategy of performing a Wick-rotation simply does not work. First, we
don't know which time coordinate to Wick-rotate. Second, even if we
just choose one and perform the rotation, generically, the resulting
metric will not be Lorentzian but {\it complex}. The overall situation
is the following.  Given an analytic Lorentzian metric, one can
complexify the manifold and extend the metric analytically. However,
the resulting complex manifold need not admit {\it any} Euclidean
section. (Conversely, we may analytically continue an Euclidean metric
and the resulting complex space-time need not have any Lorentzian
section.) This is not just an esoteric, technical problem. Even the
Lorentzian Kerr metric, which is {\it stationary}, does not admit an
Euclidean section. Thus, even if one did manage to solve the highly
non-trivial problem of actually constructing an Euclidean theory using
the hints provided by the computer work, without a brand new idea we
would still not be able to answer physical questions that refer to the
Lorentzian world. More generally --i.e., going beyond computational
physics-- one can hope to use the Euclidean techniques for some
specific calculations tailored to suitable approximations. However,
for the {\it construction} of a consistent quantum field theory, as of
now, the Euclidean techniques seem to be of little help.

Why not then try canonical quantization? This method lacks manifest
covariance. Nonetheless, as we will see, one can construct a
Hamiltonian framework without having to introduce any background
fields, thereby respecting the diffeomorphism invariance of the
theory. However, from the perspective of mathematical physicists, the
structure of the resulting framework has an unusual feature which one
has never seen in any of the familiar field theories: it is a
dynamically constrained system.  That is, most of the non-trivial
content of the theory lies in its constraints.  To see how this comes
about, let us first consider Yang-Mills theory in Minkowski space.  As
is well-known, the Hamiltonian description of this theory has a
constraint --the Gauss law. It tells us that, only those points
$(A_a^i(\vec{x}), E^a_i(\vec{x}))$ of the phase space represent {\it
physical} states of the classical theory for which the electric field
$E^a_i$ has zero covariant divergence, i.e., $(A_a^i, E^a_i)$
satisfies the constraint $D_aE^a_i = 0$. The canonical transformations
generated by this functional corresponds precisely to gauge
transformations on $(A_a^i, E^a_i)$.  In quantum theory, the
corresponding operator equation is imposed on wave functionals
$\Psi(A)$ to select the {\it physical} states: A state is physical if
and only if $\hat{D}_a\hat{E}^a\circ \Psi(A) =0$. This operator
equation tells us simply that the physical states are gauge invariant:
$\Psi(A) = \Psi(A^g)$, where $A^g$ is the transform of $A$ under a
local gauge transformation $g(\vec{x})$. The dynamics on these states
is generated by the Hamiltonian operator which, being gauge invariant,
maps physical states to physical states. Let us return to general
relativity.  Now, the group of space-time diffeomorphisms is the
``gauge group'' of the theory. Hence, in the Hamiltonian framework,
there are four constraints, three corresponding to ``spatial''
diffeomorphism on the 3-manifold fixed in the construction of the
phase space and one corresponding to diffeomorphisms in ``time-like
directions'' transverse to this surface. The canonical transformation
generated by this last constraint functional defines the dynamics of
the theory; it is therefore called the {\it Hamiltonian constraint}.
Thus, we have a peculiar situation: on physical classical states, the
generator of dynamics --the Hamiltonian-- vanishes identically!%
\footnote{${}^\dagger$}
{For simplicity I am assuming that the spatial slices
are compact. In the asymptotically flat case, the Hamiltonian
generating dynamics does not vanish identically; it equals a surface
term. The total energy of the theory is thus analogous to the charge
integral in QED. This situation is generic to diffeomorphism invariant
theories; it is not restricted to the Einstein-Hilbert action.}
Suddenly, then, mathematical physicists find themselves in an
unfamiliar territory and all the experience gained from canonical
quantization of field theories in 2 or 3 dimensions begins to look not
so relevant.  Normally, the key problem is that of finding a suitable
representation of the CCRs --or, an appropriate measure on the space
of states-- which lets the Hamiltonian be self-adjoint. Now, the
Hamiltonian seems trivial but all the difficulty seems concentrated in
the quantum constraints. Furthermore, while the representation of the
CCRs is to be chosen {\it prior} to the imposition of constraints, the
scalar product on the space of states has physical meaning only {\it
after} the constraints are solved. One thus needs to develop new
strategies and modify the familiar quantization program appropriately.

Finally, one may imagine using techniques from geometrical
quantization (see, e.g., Woodhouse 1980). However, since the key
difficulty lies in the constraints of the theory, a ``correct''
polarization would be the one which is preserved, in an appropriate
sense, by the Hamiltonian flows generated by the constraint functions
on the phase space (Ashtekar \& Stillermann, 1986). The problem of
finding such a polarization is closely related to that of obtaining a
general solution to Einstein's equation and therefore seems hopelessly
difficult in 4 dimensions.  (Incidentally, in 3 dimensions, the
strategy does work but only because there are no local degrees of
freedom; every solution to the field equations is flat.) One may
imagine using instead the Dirac (1964) approach to quantization of
constrained systems: as in Yang-Mills theory, use the operator
constraints to select physical states. This is in fact the procedure
one uses in familiar examples such as a free relativistic particle:
the classical constraint, $P^\a P_\a + \mu^2 = 0$ provides us,
via the Dirac strategy, the Klein-Gordon equation $\eta^{\a\b}\d_\a
\d_\b \Phi -\mu^2\Phi = 0$ which incorporates quantum dynamics.
Indeed, in the next two sections, we will use this strategy.  We will
see however that the representation best suited for solving the
quantum constraints in this framework does not arise from any
polarization what so ever on the phase space. Thus, unfortunately,
geometric quantization techniques do not seem to be well suited to
this problem.  Finally, one might imagine group theoretic method of
quantization. An appropriate canonical group was in fact found (Isham
\& Kakas, 1984a,b) and it does ``interact'' well with the constraints
of the theory which generate spatial diffeomorphisms (and triad
rotations).  However, the problem of incorporating the Hamiltonian
constraint which generates dynamics again seems hopelessly difficult.

As the discussion suggests, one needs a new quantization program that
can handle the peculiarities of general relativity. Such a program
does exist. The basic ideas were introduced by Dirac already in the
sixties and have been refined over the years by many authors. The
approach to quantum general relativity that I will discuss in the next
two sections is based on the framework developed in Ashtekar (1991)
(chapter 10; see also Ashtekar \& Tate 1993).

\bigskip\bigskip
\goodbreak
{\bf 4. A mathematical framework for quantum general relativity}

In this section, I will present a number of recent results that
provide a mathematical framework for non-perturbative, canonical
quantization of general relativity in 3 and 4 space-time dimensions.
This is {\it not} an exhaustive treatment of canonical quantum
gravity.  I will focus only on a few topics and omit most of the
details. I will, however, provide references where these can be found.
My aim is to illustrate the type of results that have been obtained
and provide a feel of the current status of the field from the
perspective of mathematical physics.

This section is divided in to three parts. In the first, I will
collect some basic results that relate general relativity to theories
of connections. In the second --and the main part-- I will report on
some recent work on calculus on the space $\AG$ of connections modulo
gauge transformations. In the third part, I will indicate how these
results are now being used in quantum gravity.

\bigskip\goodbreak
{\sl 4.1 Preliminaries}

As I mentioned in the Introduction, it is possible to regard general
relativity as a dynamical theory of connections. In 3 space-time
dimensions, the connection in question is simply the spin-connection
which enables one to transport the $SU(1,1)$ spinors along curves. In
4 dimensions, the appropriate connection turns out to be {\it chiral};
it enables one to transport (say) left-handed spinors along curves. In
both cases, one can construct a Hamiltonian formulation in which all
the equations are low order polynomials in the connection and its
``canonically conjugate variable'' --the analog of the electric field
of Yang-Mills theory.  However, unlike in the Yang-Mills theory, this
field, $E^a_i$, has a dual interpretation: it can be regarded as a
square root of the spatial metric. (In the 4-dimensional theory, this
is just a spatial triad.)  This dual interpretation enables one to
pass back and forth between Yang-Mills theory and general relativity.
In particular, the effect of Yang-Mills gauge transformations on the
electric field can be re-interpreted as a triad rotation, which leaves
the metric invariant --it is thus a gauge transformation also from the
perspective of general relativity. Consequently, the field $E^a_i$ is
constrained to be divergence-free also in general relativity. However,
as noted in section 3, general relativity has additional constraints,
a vectorial constraint that implies that physical states should be
invariant under spatial diffeomorphisms and a scalar (or Hamiltonian)
constraint which encodes dynamics of the theory.

Thus, in this connection-dynamics formulation of general relativity,
the phase space, to begin with, is the same as in the Yang-Mills
theory.  Furthermore, the constraint sub-manifold of the phase space
in general relativity is embedded in the constraint sub-manifold of
the Yang-Mills theory. The difference lies, of course, in dynamics. In
the Yang-Mills theory, it is generated by the Hamiltonian,
$$H(A,E)= \lint d^3x (E^a_i E_a^i + B^a_iB_a^i)\ , \eqno(4.1)$$
which is (gauge invariant but otherwise) unrelated to the constraint
(i.e., the Gauss law). In general relativity, there are additional
constraints and time-evolution is coded in one of them. Hence, if one
can solve the quantum constraints, one has essentially tackled the
issue of dynamics. We will not need the explicit form of the
constraints in what follows.

Let us conclude the preliminaries by noting that a key simplification
arises in 3 dimensions: the general relativity constraints imply that
the connection must be flat. Thus, in this case, the essence of the
entire theory is contained in just two constraints: the first is the
Gauss law which ensures gauge invariance and the second is flatness
which implies that the $SU(1,1)$ connection --which serves as the
configuration variable-- must be flat. In the classical theory, the
two constraints tell us that all dynamical trajectories correspond to
flat metrics; there are no local degrees of freedom, no gravity waves.
In the quantum theory, the imposition of these constraints implies
that the physical states $\Psi(A)$ are gauge invariant functions of
{\it flat} connections --i.e., are functions on the moduli space of
flat connections on the given spatial, 2-manifolds.  Since these
spaces are {\it finite} dimensional, we have quantum {\it mechanics}
rather than quantum field theory. In 4 dimensions, of course, no such
simplification occurs. We again have the general relativity
constraints in addition to the Gauss law. However, these do {\it not}
imply that the connections must be flat. In the classical theory,
there are local degrees of freedom, gravity waves and rich, albeit
complicated, dynamics. In the quantum theory, we have a genuine field
theory with infinite number of degrees of freedom.

For details, see, e.g., Ashtekar (1987, 1991) and Romano (1993).

\bigskip\goodbreak
{\sl 4.2 Integration on the space} $\AG$

As we saw in sections 1 and 3, any attempt at non-perturbative
quantization of gravity faces a host of conceptual problems.
3-dimensional general relativity is an excellent toy model to see how
these problems can be faced since it has been solved exactly (see,
e.g., Witten (1988), Ashtekar et al (1989), Nelson \& Regge (1989).)
{\it However, one needs a new mathematical framework and a variety of
new techniques.} Indeed, even the finished theory contains a number of
unfamiliar notions. The basic observables are essentially non-local;
the familiar operator-valued distributions are notably absent. To begin
with, there is no Hamiltonian; indeed, no time to evolve anything in.
There is no underlying space-time and hence no microcausality.  Yet,
the Hilbert space is well-defined and the observables are represented
by self-adjoint observables. One can, if one wishes, regard a suitable
dynamical variable as time and see how states ``evolve'' relative
to this ``internal clock.'' One can show that this evolution is
unitary.  There are physical predictions. The theory is in a from that
is unfamiliar from, say, constructive quantum field theory but has as
much physical content as one can hope for. Therefore, in my talk I
discussed this case in some detail. In this report, however, I will
forego that discussion since there are a number of reviews on the
subject (see, e.g., Carlip (1990, 1993) and Ashtekar (1991), chapter 17)
Instead, I will focus here on recent developments which encompass the
4-dimensional theory.

A key mathematical problem is to develop integration theory on the
space $\AG$ of connections modulo gauge transformations since,
heuristically, this is the domain space of quantum states.  In the
3-dimensional case, this problem is easy to solve because the
(appropriate components of the) moduli space of {\it flat} $SU(1,1)$
connections can be naturally given the structure of a finite
dimensional symplectic manifold (see e.g. Ashtekar(1991), chapter 17);
one can simply use the Liouville volume element to perform
integration. Thus, although the domain space of quantum states is
non-linear, the integration theory is simple because the space is {\it
finite} dimensional.  In 4 space-time dimensions, the situation is
again simple for the case of {\it linearized gravity} --the theory of
free gravitons in Minkowski space.  This theory can be cast in the
language of connections (see, e.g., Ashtekar (1991), chapter 11).
Integration theory is again well-developed; the domain space is now
linear and one can simply use the Gaussian measure as in free field
theories in Minkowski space.  Thus, in this case, in spite of the
presence of an infinite number of degrees of freedom, the integration
theory is straightforward because of the underlying {\it linearity}.
In the case of full, non-linear general relativity --and also for
Yang-Mills theory-- in 4-dimensions the problem is significantly more
difficult because $\AG$ is {\it both} non-linear {\it and} infinite
dimensional. Fortunately, a rigorous approach has now become available
to tackle this problem. In particular, a diffeomorphism invariant
measure has been found. I will now outline these developments and
indicate in the next sub-section how they can be used in quantum
gravity.

Fix an analytic 3-manifold $\S$ which will represent a Cauchy surface
in space-times to be considered. We will consider $SU(2)$ connections
on $\S$.  Since any $SU(2)$ bundle over a 3-manifold is trivial, we
can represent any connection by a Lie-algebra valued 1-form $A_a^i$ on
$\S$, where $a$ is the spatial index and $i$, the internal%
\footnote{${}^\dagger$}
{Results reported in this section for which explicit references are
not provided are all taken from Ashtekar and Isham (1992) and Ashtekar
and Lewandowski (1993a,b). The last two papers and those by Baez
(1993a,b) contain significant generalizations which include allowing
more general gauge groups, allowing the manifold $\S$ to be of
arbitrary dimension and allowing the connections to live in
non-trivial bundles.}.
Denote by $\A$ the space of smooth (say $C^2$) $SU(2)$-connections
equipped with one of the standard (Sobolev) topologies (see, e.g.
Mitter \& Viallet (1981)). $\A$ has the structure of an affine space.
However, what is of direct interest to us is the space $\AG$ obtained
by taking the quotient of $\A$ by $(C^3)$ local gauge transformations.
In this projection, the affine structure is lost; $\AG$ is a genuinely
non-linear space with complicated topology. To define the integration
theory, we will begin by constructing a sub-algebra of the Abelian
$C^\star$-algebra of bounded functions on $\AG$ and the desired
measures will arise from positive linear functionals on this \C. Given
any closed loop $\a$ on the 3-manifold, we can define the Wilson-loop
functional $T_\a$ on $\AG$:
$$T_\a(A):= \half \Tr\ {\cal P}\ \exp\ G\oint_\a A.dl, \eqno(4.2)$$
where the trace is taken in the fundamental representation of $SU(2)$
and the Newton's constant $G$ appears because, in general relativity,
it is $GA_a^i$ that has the dimensions of a connection; in gauge
theories, of course, this factor would be absent.%
\footnote{${}^\ddagger$}
{The letter $T$ stands for trace. We will later define other
$T$-variables which have the information about the ``electric field''
$E^a_i$ as well. Although defined on $\A$, being gauge invariant,
functions $T_\a$ project down to $\AG$ unambiguously.}
For technical reasons, we will have to restrict ourselves to piecewise
analytic loops $\a$. (This is why we needed $\S$ to be analytic. Note
that the loops need not be smooth; they can have kinks and
intersections but only at a finite number of points.) It turns out
that, due to $SU(2)$ trace identities, product of any two Wilson-loop
functionals can be expressed as a sum of other Wilson loop
functionals. Therefore, the vector space generated by finite
complex-linear combinations of these functions has the structure
of a $\star$-algebra. The functionals $T_\a$ are all bounded (between
$-1$ and $1$). Hence, the sup-norm (over $\AG$) is well-defined and we
can take the completion to obtain a \C. We will call it the {\it
holonomy \C} and denote it by $\HA$. Elements of $\HA$ are to be
thought of as the configuration variables of the theory.

Since $\HA$ is an Abelian $C^\star$-algebra with identity, we can
apply the Gel'fand theory and conclude that $\HA$ is isomorphic with
the $C^\star$-algebra of all continuous functions on a compact,
Hausdorff space $sp(\HA)$, the spectrum of the given \C $\HA$.
Furthermore, since elements of $\HA$ suffice to separate points of
$\AG$, it follows that $\AG$ is {\it densely embedded in} $sp(\HA)$.
To emphasize this point, from now on, we will denote the spectrum by
$\AGbar$ and regard it as a completion of $\AG$ (in the Gel'fand
topology). Integration theory will be defined on $\AGbar$. This is in
accordance with the common occurrence in quantum field theory: while
the classical configuration (or phase) space may contain only smooth
fields (typically taken to belong to be the Schwartz space), the
domain space of quantum states is a completion of this space in an
appropriate topology (the space of distributions).

A key difficulty with the use of the Gel'fand theory is that one
generally has relatively little control on the structure of the
spectrum. In the present case, however, we are more fortunate: a
simple and complete characterization of the spectrum is available.  To
present it, I first need to introduce a key definition. Fix a base
point $x_o$ in the 3-manifold $\S$ and regard two (piecewise analytic)
closed loops $\a$ and $\a'$ to be equivalent if the holonomy of any
connection in $\A$, evaluated at $x_o$, around $\a$ is the same as
that around $\a'$. We will call each equivalence class (a
holonomically equivalent loop or) a {\it hoop} and denote the hoop to
which a loop $\a$ belongs by $\at$. For example, $\a$ and $\a'$ define
the same hoop if they differ by a reparametrization or by a line
segment which is immediately re-traced%
\footnote{${}^\dagger$}
{For piecewise analytic loops and $SU(n)$ connections, these two are
the most general operations; two loops define the same hoop if and
only if they are related by a combination of reparametrizations and
retracings.}.
The set of hoops has, naturally, the structure of a group. We will
call it the {\it hoop group} and denote it by $\HG$. In terms of this
group, we can now present a simple characterization of the Gel'fand
spectrum $\AGbar$:
{\smallskip\narrower{\sl \noindent Every homomorphism $\hat{H}$ from
the hoop group $\HG$ to the gauge group $SU(2)$ defines an element
$\Ab$ of the spectrum $\AGbar$ and every $\Ab$ in the spectrum defines
a homomorphism $\hat{H}$ such that $\Ab(\at) = \textstyle {1\over 2}
\Tr\ \hat{H}(\at)$. This is a 1-1 correspondence modulo the trivial
ambiguity that homomorphisms $\hat{H}$ and $g^{-1}\cdot \hat{H}\cdot
g$ define the same element $\Ab$ of the spectrum.}\smallskip}
\noindent Clearly, every regular connection $A$ in $\A$ defines the
desired homomorphism simply through the holonomy operation:
$\hat{H}(\at) := {\cal P}\ \exp G\oint_\a A.dl$, where $\a$ is any
loop in the hoop $\at$. However, there are many homomorphisms which do
not arise from smooth connections. This leads to ``generalized
connections'' --i.e. elements in $\AGbar -\AG$. In particular, there
exist $\Ab$ in $\AGbar$ which have support at a single point and are
thus ``distributional. Note that this characterization of the spectrum
$\AGbar$ is {\it completely algebraic}; there is no continuity
assumption on the homomorphisms.  This property makes the
characterization very useful in practice.

 From the general representation theory of \C s, it follows that positive
linear functionals on $\HA$ are in 1-1 correspondence with regular
measures on (the compact Hausdorff space) $\AGbar$. It turns out that
the positive linear functions, in turn, are determined completely by
certain ``generating functionals'' $\Gamma(\a)$ on the space $\L$ of
loops based at $x_o$:
{\smallskip\narrower{\sl \noindent There is a 1-1 correspondence
between positive linear functionals on $\HA$ (and hence regular
measures on $\AGbar$) and functional $\Gamma(\a)$ on $\L$ satisfying:
\hfill\break
\noindent i) $\sum_i a_i T_{a_i} = 0 \implies \sum_i a_i \Gamma(\a_i)
= 0$; and, \hfill\break
\noindent ii) $\sum_{i,j}\ovr{a}_{i} a_j(\Gamma(\a_i\circ\a_j)+
\Gamma(\a_i \circ\a_j^{-1} )\ge 0$.\hfill\break
\noindent for all loops $\a_i$ and complex numbers $a_i$.}
\medskip}
\noindent The first condition implies that the functional $\Gamma$ is
well-defined on hoops. Hence we could have taken it to be a functional
on $\HG$ from the beginning. Thus, we see that there is a nice
``non-linear duality'' between the spectrum $\AGbar$ and the hoop
group $\HG$: Elements of $\AGbar$ are homomorphisms from $\HG$ to
$SU(2)$ and regular measures on $\AGbar$ correspond to certain
functionals on $\HG$. Finally, if one is interested in measures on
$\AGbar$ which are invariant under the (induced) action of
diffeomorphisms on $\S$, one is led to seek functionals $\Gamma(\a)$
which depend not on the individual loops $\a$ but rather on the
(generalized) knot class to which $\a$ belongs. (The qualification
``generalized'' refers to the fact that here we are considered loops
which can have kinks, overlaps and self-intersections. Until recently,
knot theorists considered only smoothly embedded loops.) Thus, there
is an interesting --and potentially powerful-- interplay between knot
theory and representations of the holonomy algebra $\HA$ in which the
diffeomorphism group of $\S$ is unitarily implemented.

Finally, we can make the integration theory more explicit. Consider a
subgroup $S_n$ of the hoop group $\HG$ which is generated by n
(independent) hoops. We can introduce the following equivalence
relation on $\AGbar$: $\Ab\equiv \Ab'$ if and only if their action on
all elements of $S_n$ coincides, i.e., if and only if $\Ab(\at) =
g^{-1}\cdot\Ab'(\at)\cdot g$ for all $\at \in S_n$ and some (hoop
independent) $g\in SU(2)$. {\it It turns out that the quotient space
is isomorphic to} $\SU$. Therefore, we can introduce a
notion of cylindrical functions on $\AGbar$: A function $f$ on
$\AGbar$ will be said to be {\it cylindrical} if it is the pull-back
to $\AGbar$ of a smooth function $\tw{f}$ on $\SU$ for some sub-group
$S_n$ of the hoop group. Finally, we can define integrals of these
functions $f$ on $\AGbar$ through their integrals on $\SU$ , provided
of course we equip $\SU$ with suitable measures $d\mu_n$ for each
$n$. We can then define a positive linear functional $\Gamma'$ on
on the space of cylindrical functions $f$ via:
$$ \Gamma'(f) := \int_{\SU} \tw{f} d\mu_n\ .
\eqno(4.3)$$
For the functional to be well-defined, of course, the family of
measures $d\mu_n$ on $\SU$ must satisfy certain consistency
conditions.  It turns out that these requirements can be met and,
furthermore, the resulting functionals $\Gamma'(f)$ define regular
measures on $\AGbar$.  A particularly natural choice (and, not
surprisingly, the first to be discovered) is to let $d\mu_n$ be simply
induced on $\SU$ by the Haar-measure on $SU(2)$. We then have the
following results: {\smallskip\narrower{\sl
\noindent i)The consistency conditions are satisfied; the left side of
(4.3) is well-defined for all cylindrical functions $f$ on $\AGbar$;
\hfill\break
\noindent ii) The generalized holonomies $T_{\at}$ are cylindrical
functionals on $\AGbar$ and $\Gamma(\a): = \Gamma' (T_{\at})$ defined
via (4.3) serves as a generating functional for a faithful, cyclic
representation of the honomony $C^\star$-algebra $\HA$ which ensures
that $d\mu$ is a regular, strictly positive measure on $\AGbar$;
\hfill\break
\noindent iii) The measure $d\mu$ is invariant under the induced action
of the diffeomorphism group on $\S$.}\smallskip}
(The knot invariant defined by $d\mu$ is a genuinely generalized one;
roughly, it counts the number of self-overlaps in any given loop.)

This measure is in some ways analogous to the Gaussian measure on
linear vector spaces. Both can be obtained by a ``cylindrical
construction.'' The Gaussian measure uses the natural metric on
$\real^n$ while the above measure uses the natural (induced) Haar
measure on $\SU$. They are both regular and strictly positive. This
leads us to ask if other properties of the Gaussian measure are
shared.  For instance, we know that the Gaussian measure is
concentrated on distributions; although the smooth fields are dense in
the space of distributions in an appropriate topology, they are
contained in a set whose total measure is zero. Is the situation
similar here? The answer turns out to be affirmative. The classical
configuration space $\AG$ with which we began is dense in the domain
space $\AGbar$ of quantum states in the Gel'fand topology.  However,
$\AG$ is contained in a set whose total measure is zero. The measure
is again concentrated on ``generalized'' connections in $\AGbar$
(Marolf \& Mour\~ao (1993)).  In a certain sense, just as the Gaussian
measures on linear spaces originate in the harmonic oscillator, the
new measure on $\AGbar$ originates in a (generalized) rotor (whose
configuration space is the $SU(2)$ group-manifold).  However, the
measure is, so to say, ``genuinely'' tailored to the underlying
non-linearity. It is {\it not} obtained by ``perturbing'' the Gaussian
measure.

With the measure $d\mu$ at hand, we can consider the Hilbert space
$L^2(\AGbar, d\mu)$ and introduce operators on it. This is {\it not}
the Hilbert space of physical states of quantum gravity since we have
not imposed constraints. It is a fiducial, kinematical space which
enables us to regularize various operators (in particular, the quantum
constraint operators). The configuration operators are associated with
the generalized Wilson loop functionals: $\hat{T}_{\at}\circ\Psi(\Ab)
= \Ab(\at)\Psi(\Ab)$. One can show that there are bounded,
self-adjoint operators on the Hilbert space. There are also ``momentum
operators'' --associated with closed, 2-dimensional ribbons or strips
in the 3-manifold $\S$-- which are gauge invariant and linear in the
electric field. One can show that these are also self-adjoint (but
unbounded). Finally, since $d\mu$ is invariant under the induced
action of the diffeomorphism group of $\S$, this group acts unitarily.

The next task is to represent quantum constraints as well-defined
operators on the Hilbert space and then solve them, i.e., find their
kernels. The first step has been completed for the three vector
constraints of general relativity. As for the kernel, typically, the
constraint operators are self-adjoint on the kinematical Hilbert space
and zero is in the continuous part of their spectrum. The physical
states --elements of the kernel-- are thus {\it not} normalizable;
they do not belong to the Hilbert space. Rather, they belong to the
(appropriately constructed) rigged Hilbert spaces (Haji\v cek, 1993).
Consider for example, the simple case of a free relativistic particle,
where the classical constraint is $P^{\a} P_{\a} +\mu^2 =0$. In this
case, the kinematical Hilbert space can be taken to be $L^2(\real^4)$.
This space is needed to translate the classical constraint function to
a well-defined operator (whose kernel can then be found). The
operator, of course, is $\eta^{\a\b}\d_\a\d_\b -\mu^2$.  No (non-zero)
element in its kernel is normalizable in $L^2(\real^4)$.  These
elements belong to the rigged Hilbert space; in the momentum space,
they are distributions with support on the mass shell. One wishes to
carry out a similar construction in the gravitational case.  For this,
one needs to introduce the appropriate rigged Hilbert spaces. This is
an open problem where input from the mathematical physics community
would be most useful. To summarize, the present status is that the
operators representing the vector constraints of general relativity
are well-defined and self-adjoint on $L^2(\AGbar, d\mu)$.  As we will
see in the next sub-section, it is intuitively clear what their kernel
is. What is needed is a precise, rigorous result.  For the Hamiltonian
constraint, we are yet to show that the operator is well-defined and
all considerations are heuristic at this stage.

Thus, the integration theory based on the measure $d\mu$ is being used
as the mathematical basis in the quantization of general relativity in
the connection-dynamics approach.  However, these techniques may be
used also in other theories of connections which are diffeomorphism
invariant and perhaps even in Yang-Mills theory which is not
diffeomorphism invariant. We saw that the full domain space of quantum
theory, $\AGbar$, can be thought of as the space of homomorphisms from
the full hoop group $\HG$ to the gauge group $SU(2)$. Given a finitely
generated sub-group $S_n$ of the hoop group, we can consider the space
of homomorphisms from it to $SU(2)$. This provides the space $\SU$
which is precisely the domain space of quantum states of a lattice
gauge theory where the lattice is not rectangular but tailored to the
given subgroup $S_n$ of the gauge group.  Thus, what we have is a set
of ``floating lattices,'' each associated with a finitely generated
subgroup of the hoop group. The space $\AGbar$ can be rigorously
recovered as a projective limit of the configuration spaces of lattice
theories (Marolf \& Mour\~ao 1993).  This construction is potentially
quite powerful; it may enable one to take continuum limits of
operators of lattice theories in a completely new fashion. The limit
is obtained not by taking the lattice separation to zero but by
enlarging lattices to probe the continuum connections better and
better, i.e., by considering larger and larger subgroups of the hoop
group.

\bigskip\goodbreak
{\sl 4.3 Loop representation of quantum gravity}

In non-perturbative quantum gravity, to date, most progress has been
made in the so-called ``loop representation'' in which {\it quantum
states arise as functionals of closed loops on the 3-manifold}.  This
representation can also be used in, e.g., Yang-Mills theories (Loll
1992) and has led to concrete results in the lattice formulations
(Br\"ugmann 1991). However, the representation is particularly
well-suited in the gravitational case since it seems to be best
suited for solving the quantum constraints. In particular, the
solutions to the vector (or diffeomorphism) constraints, referred to
in the last sub-section, are explicit in this representation.  In this
sub-section, I will outline various results that have been obtained so
far in this framework. From the perspective of mathematical physics,
some of the results I will report here are still heuristic.  However,
there does not seem to be any difficulty of principle in making them
rigorous. This may well be a fertile area for young researchers in
the field.

The loop representation was introduced as a heuristic device by
Rovelli and Smolin (1990) in the context of quantum gravity (and
somewhat earlier, but in a somewhat different fashion, by Gambini and
Trias (1986) in the context of Yang-Mills theory.) These ideas can be
now made rigorous using the framework outlined in the last
sub-section. Consider, as before, the Hilbert space $L^2(\AGbar,
d\mu)$ the elements of which are functionals $\Psi(\Ab)$ of
generalized connections $\Ab$. Since the generalized Wilson loop
functionals $T_{\at}(\Ab) :=\Tr \Ab(\at)$ belong to this space, we can
define the following transform:
$$\psi(\at) := \int_{\AGbar} \overline{T_{\at}}(\Ab)
\Psi(\Ab) \ d\mu\ ,\eqno(4.4)$$
to pass from functionals $\Psi(\Ab)$ of generalized connections to
functionals $\psi(\at)$ of hoops%
\footnote{${}^\dagger$}
{Since the left side is a function of hoops, we should,
strictly, use the terms hoop representation and the hoop transform.
However, in various calculations, it is often convenient to lift these
functionals from the hoop group $\HG$ to the space of loops $\L$.
Therefore, as in most of the literature on the subject, we will not
keep a careful distinction between loops and hoops in what follows.}.
This has some similarities with the Fourier transform
$$\psi(k) := \int_{-\infty}^{\infty}\ e^{ik\cdot x} \Psi(x)\ dx\
 ,\eqno(4.5)$$
that enables one to pass from the position to the momentum
representation. The role of the integral kernel, $\exp ik\cdot x$ is
now played by the generalized Wilson loop functional. The
Rovelli-Smolin transform is faithful. However, a nice characterization
of the space of functionals $\psi(\at)$ obtained through this
transform is not available.  In particular, the integration theory on
the hoop group $\HG$ has not yet been developed and therefore, at
present, there is no analog of the Plancharel theorem which makes the
Fourier transform so powerful. This is one of the open problems
looking longingly to mathematical physicists for help.

The transform as it is written above is however well-defined.
Therefore one can take operators from the connection side and write
them on the hoop side. Typically, these involve simple geometric
operations on the loop agruments --composing, breaking, and rerouting
loops. For example, the generalized Wilson-loop operators which act on
$L^2(\AGbar, d\mu)$ simply by multiplication, $\hat{T}_{\bt}\circ
\Psi(\Ab) = \Ab(\bt) \Psi(\Ab)$, can be transported to the loop states
and their action is given by:
$$(\hat{T}_{\bt}\circ \psi)(\at) = \half \big(\psi(\at\cdot\bt) +
\psi(\at\cdot\bt^{-1})\big), \eqno(4.6)$$
where $\at\cdot\bt$ is the hoop obtained by composing $\at$ and $\bt$
in the hoop group. Similarly, one can transport other operators. Of
particular interest are the vector or the diffeomorphism constraint
operators. There is one such operator $\hat{C}(V)$ associated with every
vector field $V^a$ on $\S$ and their action is given simply by:
$$\hat{C}(V)\circ\psi(\a) = \lim_{\epsilon\mapsto 0}{1\over
\epsilon}\big(\psi(\a^{V}_{\epsilon}) -\psi(\a)\big)\ ,\eqno(4.7)$$
where $\a^{V}_{\epsilon}$ is the loop obtained by displacing $\a$ along
the integral curves of $V^a$ an affine parameter distance $\epsilon$.
Thus, the regularized operator corresponding to the diffeomorphism
constraint does what one intuitively expects it to do: it drags the
loop in the argument of the wave function along the diffeomorphism.
Therefore, it is intuitively clear that the kernel of this constraint
consists of functionals of loops which remain unchanged if the loop is
replaced by a diffeomorphic one. That is, the elements of the kernel are
functions of generalized knot classes on the 3-manifold. It is remarkable
that in the loop representation, one can write down the {\it general}
solution to the three of the four constraints of general relativity in a
simple, geometrical way and the solutions relate quantum gravity with
knot theory. As I remarked earlier, one would like to make these results
rigorous by specifying the regularity conditions on the permitted knot
invariants using rigged Hilbert spaces.

The remaining (Hamiltonian) constraints are more difficult. They {\it
have been} transcribed in the loop representation and various
apparently distinct methods of doing so have led to equivalent results
(Br\"ugmann \& Pullin (1993)). Thus, there is a general feeling that
one is on the ``right track.''  However, these results are not
rigorous. A promising new direction is being pursued by Gambini and
his collaborators where the hoops group is replaced by a group of
``smoothened out'' hoops (Di Bortolo et al, 1993). This group has the
structure of a Lie group and offers a new approach to the problem of
regularization of various operators including the Hamiltonian
constraints. This approach has already led to some interesting, new
solutions to the Hamiltonian constraints which are related to some
well-known knot invariants (Br\"gmann et al, 1992a,b). This general
area of research related to the Hamiltonian constraint is one of the
centers of current activities although one seems quite far from
finding the generic solution to these constraints even at a heuristic
level.

\bigskip\bigskip
\goodbreak
{\bf 5. Waving a classical geometry with quantum threads}

I will now discuss two striking results that have emerged --already at
a kinematical level, prior to the imposition of quantum constraints--
from the loop representation (Ashtekar et al 1992). The first is that
certain operators representing {\it geometrical observables} can be
regulated in a way that respects the diffeomorphism invariance of the
underlying theory.  What is more, these regulated operators are finite
{\it without any renormalization}. Using these operators, one can ask
if there exist loop states which approximate smooth geometry at large
scales.  One normally takes for granted that the answer to such
questions would be obviously ``yes.''  However, in genuinely
non-perturbative treatments, this is by no means clear a priori; one
may be working in a sector of a theory which does not admit the
correct or unambiguous classical limit.  For example, the sector may
correspond to a confined phase which has no classical analog or the
limit may yield a wrong number even for the macroscopic dimensions of
space-time! The second main result of this subsection is that not only
is the answer to the question raised above in the affirmative but,
furthermore, {\it these states exhibit a discrete structure of a
definite type at the Planck scale}. (For further details, see, e.g.,
Rovelli \& Smolin (1990) and Ashtekar (1992), Smolin (1993).)

Let us begin with the issue of regularization. As noted in section 4,
in the present framework, the spatial metric is constructed from
products of ``electric fields'' $E^a_i$. It is a ``composite'' field
given by%
\footnote{${}^\dagger$}
{More precisely, the situation is as follows. Since there is no
background metric, the ``momenta'' $E^a_i$ are actually vector
densities of weight one, whence the composite field $q^{ab}$ is of
density weight two; it is the determinant of the covariant metric
multiplied by the contravariant metric. In what follows, these density
weights are important for the details of the arguments.  However, for
brevity, I will not dwell on this point any further.}
${q}^{ab}(x) ={E}^{ai}(x){E}^b_i(x)$. In the quantum theory,
therefore, this operator must be regulated. The obvious possibility is
point splitting. One might set $q^{ab}(x) = \lim_{y\to x} {E}^{ai}(x)
{E}^b_i(y)$.  However, the procedure violates gauge invariance since
the internal indices at two {\it different} points have been
contracted. There is, however, a suitable modification that will
ensure gauge invariance. Consider the field $T^{aa'}[\a] (y',y)$,
labelled by a closed loop $\a$ and points $y$ and $y'$ thereon,
defined in the classical theory by:
$$T^{aa'}[\a ] (y,y') := {1\over 2} \Tr\big[({\cal P} \exp\-
G\int_{y'}^y A_b dl^b) {E}^a(y')\- ({\cal P}\exp \- G\int_y^{y'}A_c
dl^c)\- {E}^{a'}(y)\big]. \eqno(5.1)$$
In the limit $\a$ shrinks to zero, $T^{aa'}[\a](y,y')$ tends to $-4
q^{aa'}$.

Now, in quantum theory, one can define the action of the operator
$\hat{T}^{aa'}[\a](y,y')$ directly on the loop states $\psi(\b)$.  The
explicit form will not be needed here. We only note that using the
bra-ket notation, $\psi(\b) = \IP{\b}{\psi}$ the action can be
specified easily. Indeed, $\bra{\b}\circ\hat{T}^{aa'}[\a](y,y')$ is
rather simple: if a loop $\beta$ does not intersect $\a$ at $y$ or
$y'$, the operator simply annihilates the bra $\langle\beta\mid$ while
if an intersection does occur, it breaks and re-routes the loop
$\beta$, each routing being assigned a specific weight. One may
therefore try to define a quantum operator $\hat{q}^{aa'}$ as a limit
of $\hat{T}^{aa'}[\a ]$ as $\a$ shrinks to zero. The resulting
operator does exist after suitable regularization {\it and}
renormalization. However (because of the density weights involved) the
operator necessarily carries a memory of the background metric used in
regularization. Thus, the idea of defining the metric operator again
fails.  In fact one can give general qualitative arguments to say that
there are no {\it local}, operators which carry the metric information
{\it and} which are independent of background fields (used in the
regularization). Thus, in quantum theory, the absence of background
fields introduces new difficulties.  That such difficulties would
arise was recognized quite early by Chris Isham and John Klauder.

There do exist, however, {\it non-local} operators which can be
regulated in a way that respects diffeomorphism invariance.

As the first example, consider the function $Q(\omega)$ --representing
the smeared 3-metric-- on the classical phase space, defined by
$$Q(\omega):= \int d^3x\ \ (q^{ab}\omega_a\omega_b)^{1\over 2}\- ,
\eqno(5.2)$$
where $\omega_a$ is any smooth 1-form of compact support%
\footnote{${}^\dagger$}
{Note that the integral is well-defined without the need of a
background volume element because $q^{ab}$ is a density of weight
two.}.
It is important to emphasize that, in spite of the notation,
$Q(\omega)$ is {\it not} obtained by smearing a {\it distribution}
with a test field; because of the square-root, $Q(\omega)$ is {\it
not} linear in $\omega$. We can, nonetheless define the corresponding
quantum operator as follows. Let us choose on $\Sigma$ test fields
$f_\epsilon (x,y)$ (which are densities of weight one in $x$ and) which
satisfy:
$$\lim_{\epsilon\to 0} \int_\Sigma \- d^3x\-\- f_\epsilon(x,y)\- g(x)
= g(y) \eqno(5.3)$$
for all smooth functions of compact support $g(x)$. If $\Sigma$ is
topologically $\real^3$, for example, we can construct these test
fields as follows:
$$f_\epsilon (x,y) = {\sqrt{h(x)}\over {\pi^{3\over 2}\epsilon^3}}
\-\- \exp -{\mid \vec x -\vec y\mid^2\over 2\epsilon^2},
\eqno(5.4)$$
where $\vec x$ are the cartesian coordinates labeling the point $x$
and $h(x)$ is a ``background'' scalar density of weight 2. Next, let
us define
$$ q^{aa'}_\epsilon(x) = -{1\over 4}\int_\Sigma d^3y \int_\Sigma
d^3y' f_\epsilon(x,y) f_\epsilon(x,y') T^{aa'}(y,y').
\eqno(5.5)$$
As $\epsilon$ tends to zero, the right side tends to $q^{ab}$ because
the test fields force both the points $y$ and $y'$ to approach $x$,
and hence the loop passing through $y, y'$, used in the definition of
$T^{aa'}(y,y')$, to zero. It is now tempting to try to define a local
metric operator $\hat{q}^{aa'}$ corresponding to $q^{aa'}$ by
replacing $T^{aa'}(y,y')$ in (5.5) by its quantum analog and then
taking the limit. One finds that the limit does exist provided we
first renormalize $\hat{q}_\epsilon^{aa'}$ by an appropriate power of
$\epsilon$. However, as before, the answer depends on the background
structure (such as the density $h(x)$) used to construct the test
fields $f_\epsilon (x,y)$. If, however, one tries to construct the
quantum analog of the {\it non-local} classical variable $Q(\omega)$,
this problem disappears. To see this, let us first express $Q(\omega)$
using (5.5) as:
$$ Q(\omega) = \lim_{\epsilon\to 0} \int_\Sigma \- d^3x \-\-
(q_\epsilon^{aa'}\omega_a\omega_{a'})^{1\over 2}.
\eqno(5.6)$$
The required quantum operator $\hat{Q}(\omega)$ on the loop states
can now be obtained by replacing $T^{aa'}(y,y')$ by the operator
$\hat{T}^{aa'}(y,y')$. A careful calculation shows that: i) the
resulting operator {\it is} well-defined on loop states; ii) no
renormalization is necessary, i.e., the limit is automatically {\it
finite}; and, iii) the final answer carries no imprint of the
background structure (such as the density $h(x)$ or, more generally,
the specific choice of the test fields $f_\epsilon (x,y)$) used in
regularization. To write out its explicit expression, let me restrict
myself to smooth loops $\a$ without any self-intersection. Then, the
action is given simply by:
$$ \langle\a\mid\circ\hat{Q}(\omega ) = \l_P^2\ \ \oint_\a ds
|\dot{\a}^a \omega_a|\>\cdot \langle \a\mid , \eqno(5.7)$$
where $l_P= \sqrt{G\hbar}$ is the Planck length, $s$, a parameter
along the loop and $\dot{\a}^a$ the tangent vector to the loop. In
this calculation, the operation of taking the square-root is
straightforward because the relevant operators are diagonal in the
loop representation.  This is analogous to the fact that, in the
position representation of non-relativistic quantum mechanics, we can
set $<x|\circ (\hat{X}^2)^{\textstyle{1\over 2}} = <x|\cdot |x|$
without recourse to the detailed spectral theory. The $G$ in $l_P$ of
(5.7) comes from the fact that $GA_a^i$ has the usual dimensions of a
connection while $\hbar$ comes from the fact that $\hat{E}^a_i$ is
$\hbar$ times a functional derivative. The final result is that, on
non-intersecting loops, the operator acts simply by multiplication:
the loop representation is well-suited to find states in which the
3-geometry --rather than its time evolution-- is sharp.

The second class of operators corresponds to the area of 2-surfaces.
Note first that, given a smooth 2-surface S in $\Sigma$, its area
${\cal A}_S$ is a function on the classical phase space. We first
express it using the classical loop variables. Let us divide the
surface $S$ into a large number $N$ of area elements $S_I, I=1,2...N$,
and set ${\cal A}_I^{\rm appr}$ to be
$$ {\cal A}_I^{\rm appr} = -{1\over 4}\left[ \int_{S_I} d^2S^{bc}(x)\-\-
\eta_{abc}\int_{{\cal S}_I} d^2S^{ b'c'} (x')\-\- \eta_{a'b'c'}\,
   T^{aa'}(x,x') \right]^{1\over 2}, \eqno(5.8)$$
where $\eta_{abc}$ is the (metric independent) Levi-Civita density of
weight $-1$. It is easy to show that ${\cal A}_I^{\rm appr}$
approximates the area function (on the phase space) defined by the
surface elements $S_I$, the approximation becoming better as $S_I$
--and hence loops with points  $x$ and $x'$ used in the definition of
$T^{aa'}$-- shrink.  Therefore, the total area ${\cal A}_S$ associated
with $S$ is given by
$$ {\cal A}_S = \lim_{N \rightarrow\infty}\, \, \sum_{I=1}^{N} \-
   {\cal A}_I^{\rm appr}. \eqno(5.9)$$
To obtain the quantum operator $\hat{\cal A}_S$, we simply replace
$T^{aa'}$ in (5.8) by the quantum loop operator $\hat{T}^{aa'}$.
This somewhat indirect procedure is necessary because, as indicated
above, there is no well-defined operator-valued distribution that
represents the metric or its area element {\it at a point}. Again, the
operator $\hat{\cal A}_S$ turns out to be finite. Its action,
evaluated on a nonintersecting loop $\a$ (for simplicity), is given
by:
$$  \langle \a | \circ \hat{\cal A}_S  =
   {l_p^2\over 2} \, \> I(S,\a )\- \cdot \langle\a | , \eqno(5.10)$$
where $I(S,\a)$ is simply the {\it unoriented} intersection number
between the 2-surface $S$ and the loop $\alpha$. (One obtains the {\it
un}oriented intersection number here and the absolute sign in the
integrand of (5.7) because of the square-root operation involved in
the definition of these operators.) Thus, in essence, ``a loop $\a$
contributes half a Planck unit of area to any surface it intersects.''

The fact that the area operator also acts simply by multiplication on
non-intersecting loops lends further support to the idea that the loop
representation is well-suited to ``diagonalize'' operators describing
the 3-geometry. Indeed, we can immediately construct a large set of
simultaneous eigenbras of the smeared metric and the area operators.
There is one, $\langle\a|$, associated to every nonintersecting loop
$\a$. Note that the corresponding eigenvalues of area are {\it
quantized} in integral multiples of $l_P^2/2$.  There are also
eigenstates associated with intersecting loops which, however, I will
not go into to since the discussion quickly becomes rather involved
technically.

With these operators on hand, we can now turn to the construction of
quantum loop states that approximate the classical metric $h_{ab}$ on
$\S$ on a scale large compared to the Planck length.  The basic idea
is to weave the classical metric out of quantum loops by spacing them
so that, on an average, precisely one line crosses {\it any} surface
element whose area, {\it as measured by the given} $h_{ab}$ is one
Planck unit. Such loop states will be called {\it weaves}. Note that
these states are not uniquely picked out since our requirement is
rather weak.  Indeed, given a weave approximating a given classical
metric, one can obtain others, approximating the same classical
metric.

Let us begin with a concrete example of such a state which will
approximate a {\it flat} metric $h_{ab}$.  To construct this state, we
proceed as follows. Using this metric, let us introduce a random
distribution of points on $\Sigma = \real^3$ with density $n$ (so that
in any given volume $V$ there are $nV(1+ {\cal O}(1/\sqrt{nV}))$
points). Center a circle of radius $a = (1/n)^{1\over 3}$ at each of
these points, with a random orientation.  We assume that $a<< L$, so
that there is a large number of (non-intersecting but, generically,
{\it linked}) loops in a macroscopic volume $L^3$. Denote the
collection of these circles by $\Delta_a$. As noted in section 4, due
to trace identities, products of Wilson loop functionals $T_{\at}$ can
be expressed as linear combinations of Wilson loop functionals. As a
consequence, it turns out that the bras defined by multi-loops are
equivalent to linear combinations of single loop bras. Therefore, for
each choice of the parameter $a$, there is a well-defined bra
$\langle\Delta_a|$. This is our candidate weave state.

Let us consider the observable $\hat{Q}[\omega]$. To see if
$\langle\Delta_a|$ reproduces the geometry determined by the classical
metric $h_{ab}$ on a scale $L>>l_p$, let us introduce a 1-form
$\omega_a$ which is {\it slowly varying on the scale} $L$ and compare
the value $Q[\omega](h)$ of the classical $Q[\omega]$ evaluated at the
metric $h_{ab}$, with the action of the quantum operator
$\hat{Q}[\omega]$ on $\langle\Delta_a|$. A detailed calculation yields:
$$ \langle\Delta_a|\circ \hat{Q}[\omega] =  \left[{\pi\over 2} \- \
({l_p\over a})^2 \, Q[w](h) + {\cal O}({a\over L})\right]\-  \cdot
\langle\Delta_a|. \eqno(5.11)$$
Thus, $\langle\Delta_a|$ is an eigenstate of $\hat{Q}[\omega ]$ and
the corresponding eigenvalue is closely related to $Q[\omega](h)$.
However, even to the leading order, the two are unequal {\it unless}
the parameter $a$ --the average distance between the centers of
loops-- {\it equals} $\sqrt{\pi/2}\, l_p$. More precisely, (5.11)
can be interpreted as follows. Let us write the leading coefficient on
the right side of this equation as $(1/4)(2\pi a/l_p)(nl_p^3)$.  Since
this has to be unity for the weave to reproduce the classical value
(to leading order), we see that $\Delta_a$ should contain, on an
average, one fourth Planck length of curve per Planck volume, where
lengths and volumes are measured using $h_{ab}$.

The situation is the same for the area operators $\hat{\cal A}_S$. Let
$S$ be a 2-surface whose extrinsic curvature varies slowly on a scale
$L >>l_P$. One can evaluate the action of the area operator on
$\langle \Delta_a|$ and compare the eigenvalue obtained with the value
of the area assigned to $S$ by the given flat metric $h_{ab}$. Again,
the eigenvalue can be re-expressed as a sum of two terms, the leading
term which has the desired form, except for an overall coefficient
which depends on the mean separation $a$ of loops constituting
$\Delta_a$, and a correction term which is of the order of ${\cal
O}({a\over L})$. We require that the coefficient be so adjusted that
the leading term agrees with the classical result. This occurs, again,
precisely when $a = \sqrt{\pi/2}\- l_p$. It is interesting to note
that the details of the calculations which enable one to express the
eigenvalues in terms of the mean separation are rather different for
the two observables. In spite of this, the final constraint on the
mean separation is {\it precisely} the same.

Let us explore the meaning and implications of these results.
\item{1)} {It is generally accepted that, to obtain
classical behavior from quantum theory, one needs two things: i) an
appropriate coarse graining, and, ii) special states. In our
procedure, the slowly varying test fields $\omega_a$ and surfaces $S$
with slowly varying extrinsic curvature enable us to perform the
appropriate coarse graining while weaves --with the precisely tuned
mean separation $a$-- are the special states. There is, however,
something rather startling: The restriction on the mean separation $a$
--i.e., on the {\it short distance} behavior of the multi-loop
$\Delta_a$-- came from the requirement that $\langle\Delta_a|$ should
approximate the classical metric $h_{ab}$ on {\it large scales} $L$!}
\item{2)} {In the limit $a\to\infty$, the eigenvalues of the two
operators on $\langle\Delta_a|$ go to zero. This is not too surprising.
Roughly, in a state represented by any loop $\a$, one expects the
quantum geometry to be excited just at the points of the loops. If the
loops are {\it very} far away from each other as measured by the
fiducial $h_{ab}$, there would be macroscopic regions devoid of
excitations where the quantum geometry would seem to correspond to a
zero metric.}
\item{3)} {The result of the opposite limit,  however, {\it is}
surprising.  One might have naively expected that the best
approximation to the classical metric would occur in the continuum
limit in which the separation $a$ between loops goes to zero. However,
the explicit calculation outlined above shows that this is not the
case: as $a$ tends to zero, the leading terms in the eigenvalues of
$\hat{Q}[\omega]$ and ${\cal A}_S$ actually diverge!  (One's first
impulse from lattice gauge theories may be to say that the limit is
divergent simply because we are not rescaling, i.e., renormalizing the
operator appropriately. Note, however, that, in contrast to the
calculations one performs in lattice theories, here, we {\it already}
have a well defined operator in the continuum. We are only probing the
properties of its eigenvectors and eigenvalues, whence there is
nothing to renormalize.)  It is, however, easy to see the reason
underlying this behavior. Intuitively, the factors of the Planck
length in (5.7) and (5.10) force each loop in the weave to
contribute a Planck unit to the eigenvalue of the two geometrical
observables.  In the limit $a\to 0$, the number of loops in any fixed
volume (relative to the fiducial $h_{ab}$) grows unboundedly and the
eigenvalue diverges.}
\item{4)} {It is important to note the structure of the argument. In
non-perturbative quantum gravity, there is no background space-time.
Hence, terms such as ``slowly varying'' or ``microscopic'' or
``macroscopic'' have, a priori, no physical meaning. One must do some
extra work, introduce some extra structure to make them meaningful.
The required structure should come from the very questions one wants
to ask. Here, the questions had to do with approximating a classical
geometry. Therefore, we could {\it begin} with classical metric
$h_{ab}$. We used it repeatedly in the construction: to introduce the
length scale $L$, to speak of ``slowly varying'' fields $\omega_a$ and
surfaces $S$, and, to construct the weave itself. The final result is
then a consistency argument: If we construct the weave according to
the given prescription, then we find that it approximates $h_{ab}$ on
macroscopic scales $L$ provided we choose the mean separation $a$ to
be $\sqrt{\pi/2} l_p$, where all lengths are measured relative to the
same $h_{ab}$.}
\item{5)} {Note that there is a considerable non-uniqueness in the
construction. As we noted already, a given 3-geometry can lead to
distinct weave states; our construction only serves to make the
existence of such states explicit. For example, there is no reason to
fix the radius $r$ of the individual loops to be $a$. For the
calculation to work, we only need to ensure that the loops are large
enough so that they are generically linked and small enough so that
the values of the slowly varying fields on each loop can be regarded
as constants plus error terms which we can afford to keep in the final
expression. Thus, it is easy to obtain a 2-parameter family of weave
states, parametrized by $r$ and $a$. The condition that the leading
order terms reproduce the classical values determined by $h_{ab}$ then
gives a relation between $r$, $a$ and $l_P$ which again implies
discreteness.  Clearly, one can further enlarge this freedom
considerably: There are a lot of eigenbras of the smeared-metric
and the area operators whose eigenvalues approximate the classical
values determined by $h_{ab}$ up to terms of the order ${\cal O}({l_p
\over L})$ since this approximation ignores Planck scale quantum
fluctuations. }
\item{6)} {Finally, I would like to emphasize that, at a conceptual
level, the important point is that the eigenvalues of $\hat{Q}[\omega
]$ and ${\cal A}[S]$ can be {\it discrete}.}

Let me conclude the discussion on weaves with two remarks. First, it
is not difficult to extend the above construction to obtain weave
states for curved metrics $g_{ab}$ which are slowly varying with
respect to a flat metric $h_{ab}$. Given such a metric, one can find a
slowly varying tensor field $t_a{}^b$, such that the metric $g_{ab}$
can be expressed as $t_a{}^c t_b{}^d h_{cd}$. Then, given a weave of
the type $\langle\Delta|$ considered above approximating $h_{ab}$, we
can ``deform'' each circle in the multi-loop $\Delta$ using $t_a{}^b$
to obtain a new weave $\langle \Delta |_{t}$ which approximates
$g_{ab}$ in the same sense as $\langle\Delta|$ approximates $h_{ab}$.
(See, also, Zegwaard (1992) for the weave corresponding to the
Schwarzschild black-hole.)  The second remark is that since the
weaves are eigenbras of the operators that capture the 3-geometry,
they do not approximate 4-geometries. To obtain a state that can
approximate Minkowski space-time, for example, one has to consider a
loop state that resembles a ``coherent state'' peaked at the weave
$\Delta_a$. In that state, neither the 3-geometry nor the
time-derivative thereof would be sharp; but they would have minimum
spreads allowed by the uncertainty principle. This issue has been
examined in detail by Iwasaki and Rovelli (1993).

Since these results are both unexpected and interesting, it is
important to probe their origin. We see no analogous results in
familiar theories. For example, the eigenvalues of the fluxes of
electric or magnetic fields are not quantized in QED nor do the
linearized analogs of our geometric operators admit discrete
eigenvalues in spin-2 gravity. Why then did we find qualitatively
different results? The technical answer is simply that the familiar
results refer to the {\it Fock representation} for photons and
gravitons while we are using a completely different representation
here. Thus, the results {\it are} tied to our specific choice of the
representation. Why do we not use Fock or Fock-like states? It is not
because we insist on working with loops rather than space-time fields
such as connections. Indeed, one {\it can} translate the Fock
representation of gravitons and photons to the loop picture. (See,
e.g., Ashtekar et al (1991) and Ashtekar \& Rovelli, (1992).)  And
then, as in the Fock space, the discrete structures of the type we
found in this section simply disappear. However, to construct these
loop representations, one must use a flat background metric and
essentially every step in the construction violates diffeomorphism
invariance. Indeed, there is simply no way to construct ``familiar,
Fock-like'' representations without spoiling the diffeomorphism
invariance. Thus, the results we found are, in a sense, a direct
consequence of our desire to carry out a genuinely non-perturbative
quantization without introducing any background structure.  However,
we do not have a uniqueness theorem singling out the measure $d\mu$
which was used to define the loop transform and hence to construct the
loop representation used here. There do exist other diffeomorphism
invariant measures which will lead to other loop representations. The
measure we have used is the most natural and the simplest among the
known ones.  Whether the results presented here depend sensitively on
the choice of the measure is not known.  Therefore, it would be highly
desirable to have a uniqueness theorem which tells us that, on
physical grounds, we should restrict ourselves to a specific (class
of) measure(s).

My overall viewpoint is that one should simultaneously proceed along
two lines: i) one should take these results as an indication that we
are on the right track and push heuristic calculations within this
general framework; and, ii) one should try to put the available
heuristic results on rigorous mathematical footing to avoid the danger
of ``wandering off'' in unsound directions.

\bigskip\bigskip
\goodbreak
{\bf 6. Conclusion}

In this article I have reported on some recent developments in
non-perturbative quantum general relativity in 4 dimensions. Due to
the space limitation, I could not go into details, and, to avoid the
discussion of certain technical points in the Hamiltonian framework of
general relativity, I have omitted a couple of important issues.
However, I hope I have given the general flavor of the type of
problems we face and the strategies that have been devised.  Many of
the problems may indeed be unfamiliar in the world of rigorous quantum
field theory since they are peculiar to general relativity.  There is
no background space-time, no obvious notion of microcausality, no
reference to the asymptotic states of scattering theory.  And yet,
significant progress could be made. In 3 space-time dimensions, in
particular, a complete, rigorous, mathematical treatment is available.

In 4 space-time dimensions, the program is far from being complete.
But as I have tried to argue in the last two sections, a number of
interesting results have emerged. The loop representation, in
particular, furnishes brand new tools to tackle problems arising from
the diffeomorphism invariance. Furthermore, the general program has
matured at least to the extent that there is a well-defined
mathematical framework in the background.  Indeed, a number of open
problems are precisely of the type that mathematical physicists can
now make major contributions to the field.  Finally, the results
obtained so far may themselves have interesting implications for
quantum field theory in general, quite outside the realm of quantum
gravity.

\bigskip\bigskip
\goodbreak

{\bf Acknowledgments}

I benefited a great deal from informal discussions with Professors
Buchholz, Emch, Fredenhagen, Gallovatti, Haag, Klauder, Wightman, and
Woronowicz during the workshop. I would like to thank Professor R. Sen
for organizing such a rewarding meeting and for his patience with all
administrative matters. This work was supported in part by the NSF
grant PHY93-96246 and by the Eberly research fund of the Pennsylvania
State University.
\bigskip\bigskip
\goodbreak
{\bf References}

\item{}{Agishtein, A. \& Migdal, A. (1992) Mod. Phys. Lett.
{\bf 7}, 1039-1061.}
\item{}{Amati, D. Ciafaloni, M. \& Veneziano, G. (1990) Nucl.
Phys. {\bf B347}, 550-590.}
\item{}{Aspinwall, P. S. (1993) Mirror symmetry, talk at the 1993
Rutherford-Appleton Meeting.}
\item{}{Ashtekar, A. (1987) Phys. Rev. {\bf D36}, 1587-1602.}
\item{}{Ashtekar, A. (1991) {\it Non-perturbative canonical
gravity}, World Scientific, Singapore.}
\item{}{Ashtekar, A. (1992) Mathematical problems of
non-perturbative quantum general relativity, SU-GP-92/11-2; to appear
in the proceedings of the 1992 Les Houches summer school, B. Julia
(ed), Elsevier.}
\item{}{Ashtekar, A. \& Stillerman, M. (1986) J. Math. Phys. {\bf 27},
1319-1330.}
\item{}{Ashtekar, A., Husain, V., Rovelli, C., Samuel, J. \&
Smolin, L. (1989) Class. Quan. Grav. {\bf 6}, L185-L193.}
\item{}{Ashtekar, A.,  Rovelli, C. \& Smolin, L. (1991) Phys. Rev.
{\bf D44}, 1740-1755.}
\item{}{Ashtekar, A. \& Isham, C. (1992) Class. Quan. Grav. {\bf 9},
1069-1100.}
\item{}{Ashtekar, A. \& Rovelli, C. (1992) Class. Quan. Grav. {\bf 91}
1121-1150.}
\item{}{Ashtekar, A., Rovelli, C. \& Smolin L. (1992) Phys. Rev.
Lett. {\bf 69} 237-240.}
\item{}{Ashtekar, A. \& Lewandowski, J. (1993a) Representation theory of
analytic holonomy $C^\star$ algebras; to appear in {\it Knots and
quantum gravity}, J. Baez (ed), Oxford University press.}
\item{} {Ashtekar A. \& Lewandowski, J. (1993b) Differential calculus
on $\AG$, in preparation.}
\item{}{Ashtekar, A. \& Tate, R. (1993) An extension of the Dirac program
for quantization of constrained systems: Examples,
Syracuse University Pre-print.}
\item{}{Baez, J. (1993a) Diffeomorphism-invariant generalized
measures on the space of connections modulo gauge transformations,
hep-th/9305045, to appear in the Proceedings of conference on quantum
topology, L. Crane and D. Yetter (eds).}
\item{}{Baez J. (1993b) Link invariants, functional integration and
holonomy algebras hep-th/9301063.}
\item{}{Br\"ugmann, B. (1991) Phys. Rev. {\bf D43}, 566-579}
\item{}{Br\"ugmann, B., Gambini, R. \& Pullin, J. (1992a) Phys. Rev. Lett.
{\bf 68}, 431-434.}
\item{}{Br\"ugmann, B., Gambini, R. \& Pullin, J. (1992b) In:
{\it XXth International conference on differential geometric methods
in physics}, S. Cotta \& A. Rocha (eds), World Scientific, Singapore.}
\item{}{Br\"ugmann, B. \& Pullin, J. (1993) Nucl. Phys. {\bf 390}
399-438.}
\item{}{Carlip, S. (1990) Phys. Rev. {\bf D42}, 2647-2654.}
\item{}{Carlip, S. (1993) Six ways to quantize (2+1)-dimensional
gravity, gr-qc/9305020.}
\item{}{Di Bortolo, R. Gambini \& Griego, J. (1992) Comm. Math. Phys.
(to appear)}
\item{}{Dirac, P. (1964) {\it Lectures on Quantum Mechanics}, Yeshiva
University Press, New-York.}
\item{}{Feynman, R.P. (1965) Nobel prize in physics award address.}
\item{}{Gambini, R. \& Trias A. (1986) Nucl. Phys. {\bf B278}, 436-448}
\item{}{Gross, D. \& Mende, P (1988) Nucl. Phys. {\bf B303}, 407-454.}
\item{}{Haji\v cek, P. (1993) Talk at the Munich international
symposium on developments and trends in gravitational physics.}
\item{}{Isham, C. \& Kakas A. (1984a) Class. Quan. Grav. {\bf 1}, 621-632.}
\item{}{Isham, C. \& Kakas A. (1984b) Class. Quan. Grav. {\bf 1}, 633-650.}
\item{}{Iwasaki, J. \& Rovelli, C. (1993) Int. J. Mod. Phys. {\bf D1}
533-558.}
\item{}{Loll, R. (1992) Nucl. Phys. {\bf B368}, 121-142.}
\item{}{Marolf D. \& Mour\~ao J. (1993) Carrier space of the
Ashtekar-Lewandowski measure, in preparation.}
\item{}{Mitter, P.K. \& Viallet C. (1981) Commun. Math. Phys. {\bf 79}
43-58.}
\item{}{Nelson, J. \& Regge, T. (1989) {\bf A4}, 2021-2030.}
\item{}{Pullin, J. (1993) Knot theory and quantum gravity in loop
space: A premier, hep-th/9301028, to appear in the proceedings of the
Vth Mexican school on particles and fields, J.L. Lucio (ed), Word
Scientific, Singapore.}
\item{}{Romano, J. (1993) Gen. Rel. Grav. {\bf 25} 759-854.}
\item{}{Rovelli, C. \& Smolin, L. (1990) Nucl. Phys. {\bf B331}, 80-152.}
\item{}{Smolin, L. (1993) In: {\it General relativity and gravitation 1992},
 R.J. Gleiser et al (eds), Institute of Physics, Bristol, pages 229-261.}
\item{}{Streater, R. \& Wightman, A. (1964) {\it PCT, spin and statistics,
and all that}, Benjamin, New York.}
\item{}{Witten, E. (1988) Nucl. Phys. {\bf B311}, 46-78.}
\item{}{Zegwaard, J. (1992) Nucl. Phys. {\bf B378}, 288-308.}

\bye